# Stratified Analysis of "Probabilities of Causation"

**Manabu Kuroki**
Systems Innovation Dept.
Osaka University
Toyonaka, Osaka, Japan
mkuroki@sigmath.es.osaka-u.ac.jp

**Zhihong Cai**
Biostatistics Dept.
Kyoto University
Sakyo-ku, Kyoto, Japan
cai@pbh.med.kyoto-u.ac.jp

## Abstract

This paper derives new bounds for the probabilities of causation defined by Pearl (2000), namely, the probability that one observed event was a necessary (or sufficient, or both) cause of another. Tian and Pearl (2000a, 2000b) showed how to bound these probabilities using information from experimental and observational studies, with minimal assumptions about the data-generating process. We derive narrower bounds using covariates measurements that might be available in the studies. In addition, we provide identifiable case under no-prevention assumption and discuss the covariate selection problem from the viewpoint of estimation accuracy. These results provides more accurate information for public policy, legal determination of responsibility and personal decision making.

## 1 Introduction

It is an important issue to evaluate the likelihood that one event was the cause of another in practical science. For example, epidemiologists are interested in the likelihood that a particular exposure is the cause of a particular disease. In order to assess this likelihood from statistical data, the probabilities of causation have been developed, which can be divided into necessary causation, sufficient causation and necessary-and-sufficient causation. These probabilities of causation are used in epidemiology, legal reasoning, artificial intelligence, policy analysis and psychology.

Pearl (2000) and Tian and Pearl (2000a, 2000b) developed formal semantics for probabilities of causation based on structural models of counterfactuals. They presented the formal definitions for probability of necessity ($PN$), probability of sufficiency ($PS$) and probability of necessity and sufficiency ($PNS$). In addition, they showed how to bound these quantities from data obtained in experimental and observational studies. Their bounds are sharp under the minimal assumptions concerning the data-generating process. In this paper, we call their bounds Tian-Pearl bounds.

When we examine many experimental and observational studies, we find that there is some extra information that we can use in order to bound the quantities of the probabilities of causation. For example, in epidemiological studies, not only the exposure and the outcome are measured, but also some covariates such as age, gender are measured. However, Tian-Pearl bounds do not provide formulas for making full use of these information. Therefore, the aim of this paper is to provide narrower bounds for probabilities of causation by using as much as information available from experimental and observational studies.

Our main idea is to use stratified analysis in order to obtain narrower bounds. When we get the statistical data about the exposure, outcome and some covariates, we first stratify the data according to the covariates, then we calculate the bounds of probabilities of causation in each stratum, finally we derive summarized bounds on the probabilities of causation. By making use of the covariate information, we can provide narrower bounds than Tian-Pearl bounds without making additional assumptions.

This paper is organized as follows. Section 2 gives some preliminary knowledge that will be used throughout the paper. In Section 3, we introduce the probabilities of causation and give a new definition of conditional probabilities of causation. Then we propose the nonparametric bounds for probabilities of causation based on stratified analysis, and investigate some properties of the proposed bounds. In addition, we provide identifiable cases under no-prevention assumption and discuss the covariate selection problem from the viewpoint of estimation accuracy. We do simulation experiments to verify our results in section 4. In section 5, we give an example to illustrate how the proposed formulas can narrow the bounds so as to provide more evidence for health care policy making. Finally, section 6 concludes this paper.

## 2 Preliminary

In this section, we introduce the potential outcome variables that will be used to define the probabilities of causation. We consider the case where an exposure variable and an outcome variable are dichotomous. We denote $X$ as an exposure variable ($x$: true; $x'$: false) and $Y$ as an outcome variable ($y$: true; $y'$: false). In addition, let $P(x,y)$ be a strictly positive joint probability of $(X,Y) = (x,y)$ and $P(y|x)$ the conditional probability of $Y = y$ given $X = x$. Similar notations are used for other distributions.

In principle, the $i$th of the $N$ subjects has both a potential outcome $Y_x(i)$ that have resulted if $X$ had been $x$, and a potential outcome $Y_{x'}(i)$ that have resulted if $X$ had been $x'$. Then $Y_x(i) - Y_{x'}(i)$ is called the unit-level causal effect (Rubin, 2005). When the $N$ subjects in the study are considered as a random sample from some population, since $Y_x(i)$ and $Y_{x'}(i)$ can be referred as the values of random variables $Y_x$ and $Y_{x'}$ respectively, the average causal effect can be defined as

$$P(y_x) - P(y_{x'}), \tag{1}$$

where $y_x$ indicates the counterfactual sentence "Variable $Y$ would have the value $y$, had $X$ been $x$". Similar notations are used for other potential outcomes.

The outcome $Y_x$ is observed only if $X$ is $x$, and $Y_{x'}$ is observed only if $X$ is $x'$. This property is called the consistency (Robins, 1989), which is formulated as follows:

$$(X = x) \Rightarrow (Y_x = Y). \tag{2}$$

Thus, when a randomized experiment is conducted and compliance is perfect, the average causal effect is

$$P(y_x|x) - P(y_{x'}|x') = P(y|x) - P(y|x'), \tag{3}$$

which is equal to equation (1). On the other hand, when a randomized experiment is difficult to conduct and only observational data is available, we can still estimate the average causal effect according to the strong-ignorable-treatment-assignment (SITA) condition (Rosenbaum and Rubin, 1983). That is, for the treatment variable $X$, if there exists such a set $S$ of observed covariates that $X$ is conditionally independent of $(Y_x, Y_{x'})$ given $S$, denoted as $X \perp\!\!\!\perp (Y_x, Y_{x'}) | S$, we shall say treatment assignment is strongly ignorable given $S$, or $S$ satisfies the SITA condition. Thus, equation (1) is estimable by using $S$

$$E_s\{P(y|x,s) - P(y|x',s)\}. \tag{4}$$

## 3 Nonparametric bounds based on stratification

### 3.1 Formulation

The $PN$ is defined as the expression

$$PN = \Pr(y'_{x'}|x,y), \tag{5}$$

which stands for the probability that event $y$ would not have occurred in the absence of $x$, given that $x$ and $y$ did in fact occur. In this paper, we define a conditional $PN$ given $S = s$ as

$$PN(s) = \Pr(y'_{x'}|x,y,s),$$

which stands for the probability that event $y$ would not have occurred in the absence of $x$, given that $x$ and $y$ did in fact occur in stratum $s$.

The $PS$ is defined as the expression $PS = P(y_x|x',y')$, which stands for the probability that event $y$ would have occurred in the presence of $x$, given that $x'$ and $y'$ did in fact occur. In addition, we define a conditional $PS$ given $S = s$ as $PS(s) = P(y_x|x',y',s)$ which stands for the probability in stratum $s$ that event $y$ would have occurred in the presence of $x$, given that $x'$ and $y'$ did in fact occur in stratum $s$. In this paper, we do not discuss the $PS$ since it has the same properties as the $PN$ by changing $(x,y)$ to $(x',y')$.

The $PNS$ is defined as the expression

$$PNS = P(y_x, y'_{x'}), \tag{6}$$

which measures both the sufficiency and necessity of $x$ to produce $y$. In this paper, the conditional $PNS$ given $S = s$ is defined as

$$PNS(s) = P(y_x, y'_{x'}|s),$$

which measures both the sufficiency and necessity of $x$ to produce $y$ in stratum $s$.

When an experimental study or an observational study is conducted, it is the usual case that not only the exposure and the outcome variables are measured, but also some information on background factors and intermediate factors are available. Then, making full use of these available information will give narrower bounds on the $PN$ and the $PNS$, which is the aim of this section. The bounds can be obtained by applying the linear programming technique (Tian and Pearl, 2000a, 2000b) to conditional causal effects (Tian, 2004). In this paper, we provide a simple proof by using equation (2) and recursive factorizations of probabilities.

Letting $S$ be a set of observed variables, when we stratify the subjects according to the levels of $S$, we can derive new bounds on the $PN$ and the $PNS$ by (i) calculating the lower and upper bounds of the conditional

$PN$ and the conditional $PNS$ within each stratum, and (ii) summarizing the lower and upper bounds in all the strata.

First, regarding the $PN$,

$$\mathrm{P}(y'_{x'}|s) = \sum_{z\in\{y,y'\}} \sum_{w\in\{x,x'\}} \mathrm{P}(y'_{x'}|w,z,s)\mathrm{P}(w,z|s), \quad (7)$$

$\mathrm{P}(y'_{x'}|x',y,s) = 0$ and $\mathrm{P}(y'_{x'}|x',y',s) = 1$ hold true from equation (2). Thus, the bounds of the conditional $PN$ given $S=s$ can be given as

$$\max\left\{\begin{array}{c}0\\ \frac{\mathrm{P}(y'_{x'}|s) - \mathrm{P}(y'|s)}{\mathrm{P}(x,y|s)}\end{array}\right\} \le PN(s)$$

$$\le \min\left\{\begin{array}{c}1\\ \frac{\mathrm{P}(y'_{x'}|s) - \mathrm{P}(x',y'|s)}{\mathrm{P}(x,y|s)}\end{array}\right\}.$$

Here, letting $PN_s$ be the $PN$ when a variable $S$ is used to evaluate it, by noting $PN_s = \sum_s PN(s)\mathrm{P}(s|x,y)$, we can derive the new bounds based on stratification

$$\sum_s \max\left\{\begin{array}{c}0\\ \frac{\mathrm{P}(y'_{x'}|s) - \mathrm{P}(y'|s)}{\mathrm{P}(x,y)}\end{array}\right\}\mathrm{P}(s) \le PN_s$$

$$\le \sum_s \min\left\{\begin{array}{c}\frac{\mathrm{P}(x,y|s)}{\mathrm{P}(x,y)}\\ \frac{\mathrm{P}(y'_{x'}|s) - \mathrm{P}(x',y'|s)}{\mathrm{P}(x,y)}\end{array}\right\}\mathrm{P}(s). \quad (8)$$

Regarding the $PNS$, it is trivial that $P(y_x, y'_{x'}|s) \le \min\{P(y_x|s), P(y'_{x'}|s)\}$ holds true. In addition, we can obtain

$$P(y_x, y'_{x'}|s) = \sum_{z\in\{y,y'\}} \sum_{w\in\{x,x'\}} P(y_x, y'_{x'}|z,w,s)P(z,w|s),$$

$$P(y_x, y'_{x'}|s) = P(y'_{x'}|s) - P(y'_x, y'_{x'}|s) = P(y'_{x'}|s)$$
$$- \sum_{z\in\{y,y'\}} \sum_{w\in\{x,x'\}} P(y'_x, y'_{x'}|z,w,s)P(z,w|s),$$

$$P(y_x, y'_{x'}|s) = P(y_x|s) - P(y_x, y_{x'}|s) = P(y_x|s)$$
$$- \sum_{z\in\{y,y'\}} \sum_{w\in\{x,x'\}} P(y_x, y_{x'}|z,w,s)P(z,w|s),$$

and

$$P(y_x, y'_{x'}|s) = P(y_x|s) - P(y_{x'}|s) + P(y'_x, y_{x'}|s)$$
$$= P(y_x|s) - P(y_{x'}|s)$$
$$+ \sum_{z\in\{y,y'\}} \sum_{w\in\{x,x'\}} P(y'_x, y_{x'}|z,w,s)P(z,w|s).$$

Thus, by using $\mathrm{P}(y'_{x'}|x',y,s) = 0$ and $\mathrm{P}(y'_{x'}|x',y',s) = 1$, the bounds of the conditional $PNS$ given $S=s$ can be given as

$$\max\left\{\begin{array}{c}0\\ P(y_x|s) - P(y|s)\\ P(y'_{x'}|s) - P(y'|s)\\ P(y_x|s) - P(y_{x'}|s)\end{array}\right\} \le PNS(s)$$

$$\le \min\left\{\begin{array}{c}P(y_x|s)\\ P(y'_{x'}|s)\\ P(x,y|s) + P(x',y'|s)\\ P(y_x|s) - P(y_{x'}|s) + P(x',y|s) + P(x,y'|s)\end{array}\right\}.$$

Thus, letting $PNS_s$ be the $PNS$ when a variable $S$ is used to evaluate it, by noting $PNS_s = \sum_s PNS(s)\mathrm{P}(s)$, we can obtain

$$\sum_s \max\left\{\begin{array}{c}0\\ P(y_x|s) - P(y|s)\\ P(y'_{x'}|s) - P(y'|s)\\ P(y_x|s) - P(y_{x'}|s)\end{array}\right\}P(s) \le PNS_s$$

$$\le \sum_s \min\left\{\begin{array}{c}P(y_x|s)\\ P(y'_{x'}|s)\\ P(x,y|s) + P(x',y'|s)\\ P(y_x|s) - P(y_{x'}|s) + P(x',y|s) + P(x,y'|s)\end{array}\right\}$$
$$\times P(s). \quad (9)$$

The proposed bounds (8) and (9) are narrower than Tian-Pearl bounds (Tian and Pearl, 2000a, 2000b). An intuitive explanation is that the proposed bounds always select the maximal value in lower bound and minimal value in upper bound within every stratum of $S$, while Tian-Pearl bounds always select a fixed one from all the probabilities across every stratum of $S$. In addition, it is obvious that the proposed bounds reduce to the Tian-Pearl bounds if there is no stratified analysis on $S$.

### 3.2 Property of the proposed bounds

An interesting property of the proposed bounds is that the observed variables $S$ need not be confounders between an exposure variable $X$ and an outcome variable $Y$. Fig.1 shows six cases that the observed variables $S$ may be. Here, $U$, $U_1$ and $U_2$ in Fig.1 indicate sets of unobserved variables. An arrow ($\rightarrow$) indicates that a variable of its tail has an effect on another variable of its head, and a bidirected arc ($\leftrightarrow$) indicates that two variables connected by the arc have an association with each other. Fig.1 (a) shows that the variables $S$ are confounders that have an effect on both $X$ and $Y$; Fig.1 (b) shows that the variables $S$ are prognostic factors that have an effect only on $Y$; Fig.1 (c) shows that the variables $S$ are intermediate variables that are affected by $X$ and have an effect on $Y$; Fig.1 (d) shows that the variables $S$ are variables that satisfy the instrumental variable conditions (e.g. Bowden and Turkington, 1984; Greenland, 2000); Fig.1 (e) shows

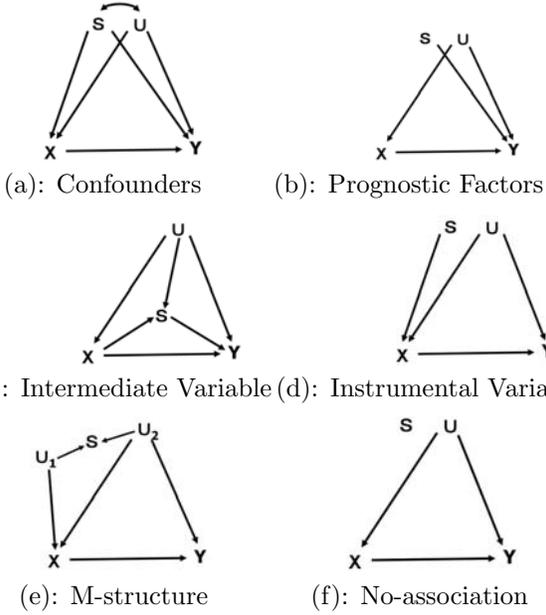

Fig.1. Six Cases of Observed Variable $S$

(a): Confounders
(b): Prognostic Factors
(c): Intermediate Variable
(d): Instrumental Variable
(e): M-structure
(f): No-association

that the covariates $S$ are in a M-structure (Greenland et al., 1999; Greenland, 2003); Fig.1 (f) shows that the variables $S$ are associated with neither $X$ nor $Y$.

When data are available from both an experimental study and an observational study, equations (8) and (9) are applicable to the former five cases, that is, we can obtain narrower bounds than Tian-Pearl bounds under these cases. Hence, when we obtain information on observed variables, they may belong to case (a), (b), (c), (d) or (e). Whichever they belong to, if we stratify the population according to the levels of $S$, we can obtain narrower bounds on probabilities of causation.

### 3.3 Identification under the "no-prevention"

In the discussion above, we proposed narrower bounds on the probabilities of causation. When the monotonicity assumption "$P(y'_x, y_{x'}) = 0$" is added, we can derive the point estimators based on covariate adjustments. In epidemiology, this assumption is often expressed as "no-prevention", which means that no individual in the population can be helped by exposure to a risk factor, that is, a hazardous exposure is either harmful or indifferent to every member of the population.

The generalization of monotonicity assumption is offered through the use of the conditional monotonicity assumption "$P(y'_x, y_{x'}|s) = 0$". This assumption implies that both $P(y'_x|y_{x'}, x', s) = P(y'_x|y, x', s) = 0$ and $P(y_{x'}|y'_x, x, s) = P(y_{x'}|y', x, s) = 0$ hold true. Then, from equation (7), by noting that $P(y'_{x'}|y', x, s) = 1 - P(y_{x'}|y', x, s) = 1$, we can obtain

$$PN(s) = \frac{P(y'_{x'}|s) - P(y'|s)}{P(x, y|s)}.$$

Thus, the $PN$ can be identified as $PN_s = \sum_s PN(s)P(s)$, which requires both observational data and experimental data. When only observational data is available, if we can observe some covariates that satisfy the SITA condition (Rosenbaum and Rubin, 1983), we can still identify the $PN$:

$$PN_s = \sum_s \frac{P(y'|x', s) - P(y'|s)}{P(x, y)} P(s). \quad (10)$$

The asymptotic variance of the $PN$ is given by

$$a.var(\hat{PN}_s)$$
$$= \sum_s \left( (1 - PN_s)^2 \frac{P(y'|x, s)P(y|x, s)}{NP(x, s)} \right.$$
$$\left. + \frac{P(y'|x', s)P(y|x', s)}{NP(x', s)} \right) \left( \frac{P(x, s)}{P(x, y)} \right)^2.$$

(Cai and Kuroki, 2005). By the similar procedure, since we can obtain $PNS(s) = P(y_x|s) - P(y_{x'}|s)$, the $PNS$ can be identified as $PNS_s = \sum_s PNS(s)P(s)$, which requires observational data or experimental data. When only observational data is available, if we can observe some covariates that satisfy the SITA condition (Rosenbaum and Rubin, 1983), we can still identify $PNS$:

$$PNS_s = \sum_s (P(y|x, s) - P(y|x', s))P(s), \quad (11)$$

which is consistent with the average causal effect provided in equation (4). The asymptotic variance of the $PNS$ is given by

$$a.var(\hat{PNS}_s)$$
$$= \sum_s \left( \frac{P(y'|x, s)P(y|x, s)}{NP(x, s)} \right.$$
$$\left. + \frac{P(y'|x', s)P(y|x', s)}{NP(x', s)} \right) P(s)^2.$$

(Cai and Kuroki, 2005).

Moreover, when two different subsets $S$ and $T$ of the covariates satisfy the SITA condition, the covariate selection problem occurs: whether it is better to use both of them than to use one of them in order to obtain a point estimator with smaller variance. Regarding this problem, we provide the following results: (1) if $Y \perp\!\!\!\perp T | \{X, S\}$ holds true from data, we can obtain

$$a.var(\hat{PN}_s) \leq a.var(\hat{PN}_{s,t}),$$

and
$$a.var(P\hat{N}S_s) \leq a.var(P\hat{N}S_{s,t});$$

(2) if $X \perp\!\!\!\perp S \mid T$ holds true from data, we can obtain

$$a.var(P\hat{N}_{s,t}) \leq a.var(P\hat{N}_t)$$

and

$$a.var(P\hat{N}S_{s,t}) \leq a.var(P\hat{N}S_t).$$

The proof is provided in Appendix.

We can describe such situations as the graph shown in Fig.2, which indicates that both $X \perp\!\!\!\perp S \mid T$ and $Y \perp\!\!\!\perp T \mid \{X, S\}$ hold true. These conditional independence relationships can be read off from the graph by the d-separation criterion (Pearl, 2000). Since both $T$ and $S$ satisfy the SITA condition relative to $(X, Y)$ and both $Y \perp\!\!\!\perp T \mid \{X, S\}$ and $X \perp\!\!\!\perp S \mid T$ hold true, we can obtain

$$a.var(P\hat{N}_s) \leq a.var(P\hat{N}_{t,s}) \leq a.var(P\hat{N}_t)$$

and

$$a.var(P\hat{N}S_s) \leq a.var(P\hat{N}S_{t,s}) \leq a.var(P\hat{N}S_t).$$

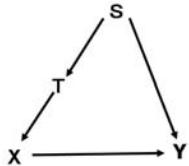

Fig. 2: Graphical representation of covariate selection

That is, the asymptotic variance when $S$ is used is smaller than that when $T$ is used under the conditions above. These results provide qualitative relationships between different covariates that satisfy the SITA condition, which indicates that it is not always better to use as much as covariate information in order to estimate the probabilities of causation.

## 4 Simulation experiment

We compare the variances in section 3.3 through simulation experiments. For simplicity, we only consider the case in Fig. 2, where there are two observed dichotomous covariates $S$ and $T$. We consider four scenarios of the odds ratios between $X$ and $T$ and between $S$ and $T$, which is shown in Table 1. The value in each cell represents the probability $P(x, s, t)$, which satisfies $X \perp\!\!\!\perp S \mid T$. Setting 1 represents the case where both the odds ratios between $X$ and $T$ and between $S$ and $T$ are larger than 1. Setting 2 represents the case where the odds ratio between $X$ and $T$ is larger than 1 but that between $S$ and $T$ is close to 1. Setting 3 represents the case where the odds ratio between $X$ and $T$ is close to 1 but that between $S$ and $T$ is larger than 1. Setting 4 represents the case where both the odds ratios between $X$ and $T$ and between $S$ and $T$ are close to 1. In addition, the setting of conditional probabilities of $Y$ given $X$ and $S$ are fixed at $(P(y|x, s_1), P(y|x, s_2), P(y|x', s_1), P(y|x', s_2)) = (0.7, 0.3, 0.8, 0.4)$. In order to verify the properties of the variances in Section 3.3, we did simulation experiments based on the four settings in sample sizes $N = 500, 1000, 1500$ and $2000$, respectively. Table 2 reports the variance estimates from 5000 replications in various sample sizes. "$S$" means the variance when a covariate $S$ is used to evaluate the $PN$ and the $PNS$, "$T$" means the variance when a covariate $T$ is used to evaluate the $PN$ and the $PNS$, and "$\{S, T\}$" means the variance when both $S$ and $T$ are used to evaluate the $PN$ and the $PNS$. The first line shows the value of variance obtained from simulation experiments, denoted as $var$, and the second line shows the value of asymptotic variances calculated from the formulas in Section 3, denoted as $a.var$.

Table 1: Four Parameter Settings

|   | Setting 1 | | | | Setting 2 | | | |
|---|---|---|---|---|---|---|---|---|
|   | $t_1$ | | $t_2$ | | $t_1$ | | $t_2$ | |
|   | $s_1$ | $s_2$ | $s_1$ | $s_2$ | $s_1$ | $s_2$ | $s_1$ | $s_2$ |
| $x$  | 0.32 | 0.08 | 0.02 | 0.08 | 0.2 | 0.05 | 0.04 | 0.16 |
| $x'$ | 0.08 | 0.02 | 0.08 | 0.32 | 0.2 | 0.05 | 0.06 | 0.24 |

|   | Setting 3 | | | | Setting 4 | | | |
|---|---|---|---|---|---|---|---|---|
|   | $t_1$ | | $t_2$ | | $t_1$ | | $t_2$ | |
|   | $s_1$ | $s_2$ | $s_1$ | $s_2$ | $s_1$ | $s_2$ | $s_1$ | $s_2$ |
| $x$  | 0.2 | 0.2 | 0.04 | 0.06 | 0.1 | 0.1 | 0.1 | 0.15 |
| $x'$ | 0.05 | 0.05 | 0.16 | 0.24 | 0.15 | 0.15 | 0.1 | 0.15 |

From Table 2, we draw the following conclusions.

(1) The ratio of variance to asymptotic variance is close to 1.0, which shows that the asymptotic variances are sufficient approximations of the variances.

(2) In all cases, the variance when $S$ is selected is smaller than the variance when $T$ or $\{S, T\}$ is selected, which is consistent with the results in section 3.3. In addition, the variance when $T$ is selected is larger than the variance when $\{S, T\}$ is selected, which is also consistent with the results in section 3.3.

(3) The variances vary in each case, which may result from the different parameter settings.

## 5 Example

The above results are applicable to analyze the data from the Northern Alberta Breast Cancer Registry (Newman, 2001). This data, which is shown in Table 3, was collected to investigate the effect of receptor level on breast cancer survival. It was also reanalyzed by Greenland (2004), with the purpose of discussing the attributable fraction and risk ratios. The size of

Table 2: Simulation Results on the Variances

| PN | | | Setting 1 | | | Setting 2 | | |
|---|---|---|---|---|---|---|---|---|
| | | | $S$ | $T$ | $\{S,T\}$ | $S$ | $T$ | $\{S,T\}$ |
| $N=500$ | | var | 0.0071 | 0.0122 | 0.0113 | 0.0081 | 0.0092 | 0.0082 |
| | | a.var | 0.0068 | 0.0120 | 0.0106 | 0.0078 | 0.0088 | 0.0078 |
| $N=1000$ | | var | 0.0035 | 0.0061 | 0.0054 | 0.0039 | 0.0044 | 0.0039 |
| | | a.var | 0.0034 | 0.0060 | 0.0053 | 0.0039 | 0.0044 | 0.0039 |
| $N=1500$ | | var | 0.0023 | 0.0040 | 0.0035 | 0.0026 | 0.0029 | 0.0026 |
| | | a.var | 0.0023 | 0.0040 | 0.0035 | 0.0026 | 0.0029 | 0.0026 |
| $N=2000$ | | var | 0.0017 | 0.0031 | 0.0027 | 0.0019 | 0.0022 | 0.0020 |
| | | a.var | 0.0017 | 0.0030 | 0.0026 | 0.0019 | 0.0022 | 0.0020 |

| PN | | | Setting 3 | | | Setting 4 | | |
|---|---|---|---|---|---|---|---|---|
| | | | $S$ | $T$ | $\{S,T\}$ | $S$ | $T$ | $\{S,T\}$ |
| $N=500$ | | var | 0.0087 | 0.0195 | 0.0168 | 0.0096 | 0.0114 | 0.0098 |
| | | a.var | 0.0083 | 0.0189 | 0.0158 | 0.0093 | 0.0111 | 0.0094 |
| $N=1000$ | | var | 0.0043 | 0.0095 | 0.0081 | 0.0048 | 0.0056 | 0.0049 |
| | | a.var | 0.0042 | 0.0094 | 0.0079 | 0.0046 | 0.0056 | 0.0047 |
| $N=1500$ | | var | 0.0028 | 0.0063 | 0.0053 | 0.0031 | 0.0037 | 0.0031 |
| | | a.var | 0.0028 | 0.0063 | 0.0053 | 0.0031 | 0.0037 | 0.0031 |
| $N=2000$ | | var | 0.0021 | 0.0048 | 0.0040 | 0.0023 | 0.0028 | 0.0024 |
| | | a.var | 0.0021 | 0.0047 | 0.0039 | 0.0023 | 0.0028 | 0.0023 |

| PNS | | | Setting 1 | | | Setting 2 | | |
|---|---|---|---|---|---|---|---|---|
| | | | $S$ | $T$ | $\{S,T\}$ | $S$ | $T$ | $\{S,T\}$ |
| $N=500$ | | var | 0.0019 | 0.0029 | 0.0026 | 0.0017 | 0.0019 | 0.0017 |
| | | a.var | 0.0018 | 0.0028 | 0.0025 | 0.0017 | 0.0019 | 0.0017 |
| $N=1000$ | | var | 0.0009 | 0.0014 | 0.0012 | 0.0008 | 0.0009 | 0.0008 |
| | | a.var | 0.0009 | 0.0014 | 0.0012 | 0.0008 | 0.0009 | 0.0008 |
| $N=1500$ | | var | 0.0006 | 0.0009 | 0.0008 | 0.0006 | 0.0006 | 0.0006 |
| | | a.var | 0.0006 | 0.0009 | 0.0008 | 0.0006 | 0.0006 | 0.0006 |
| $N=2000$ | | var | 0.0005 | 0.0007 | 0.0006 | 0.0004 | 0.0005 | 0.0004 |
| | | a.var | 0.0005 | 0.0007 | 0.0006 | 0.0004 | 0.0005 | 0.0004 |

| PNS | | | Setting 3 | | | Setting 4 | | |
|---|---|---|---|---|---|---|---|---|
| | | | $S$ | $T$ | $\{S,T\}$ | $S$ | $T$ | $\{S,T\}$ |
| $N=500$ | | var | 0.0017 | 0.0031 | 0.0027 | 0.0017 | 0.0020 | 0.0017 |
| | | a.var | 0.0017 | 0.0031 | 0.0026 | 0.0017 | 0.0020 | 0.0017 |
| $N=1000$ | | var | 0.0008 | 0.0015 | 0.0013 | 0.0009 | 0.0010 | 0.0009 |
| | | a.var | 0.0008 | 0.0015 | 0.0013 | 0.0008 | 0.0010 | 0.0008 |
| $N=1500$ | | var | 0.0006 | 0.0010 | 0.0009 | 0.0006 | 0.0007 | 0.0006 |
| | | a.var | 0.0006 | 0.0010 | 0.0009 | 0.0006 | 0.0007 | 0.0006 |
| $N=2000$ | | var | 0.0004 | 0.0008 | 0.0007 | 0.0004 | 0.0005 | 0.0004 |
| | | a.var | 0.0004 | 0.0008 | 0.0006 | 0.0004 | 0.0005 | 0.0004 |

the sample is 192 and the variables of interest are the following:

$X$: Receptor Level ($x$: high; $x'$: low).

$Y$: Survival Indicator ($y$: death; $y'$: survive).

$S$: Stage at Diagnosis ($s_1$: stage 1; $s_2$: stage 2; $s_3$: stage 3).

Table 3: Receptor Level-Breast Cancer Study (Newman, 2001)

| | $s_1$ | | $s_2$ | | $s_3$ | |
|---|---|---|---|---|---|---|
| | $x'$ | $x$ | $x'$ | $x$ | $x'$ | $x$ |
| $y$ | 2 | 5 | 9 | 17 | 12 | 9 |
| $y'$ | 10 | 50 | 13 | 57 | 2 | 6 |

In this example, we assume that $S$ satisfies the SITA condition. Since only observational data is available, the proposed bounds of equation (8) and (9) can be written as

$$\sum_s \max \left\{ \begin{array}{c} 0 \\ \dfrac{P(y'|x',s) - P(y'|s)}{P(x,y)} \end{array} \right\} P(s) \leq PN_s$$

$$\leq \sum_s \min \left\{ \begin{array}{c} \dfrac{P(x,y|s)}{P(x,y)} \\ \dfrac{P(y'|x',s) - P(x',y'|s)}{P(x,y)} \end{array} \right\} P(s)$$

and

$$\sum_s \max \left\{ \begin{array}{c} 0 \\ P(y|x,s) - P(y|x',s) \end{array} \right\} P(s) \leq PNS_s$$

$$\leq \sum_s \min \left\{ \begin{array}{c} P(y|x,s) \\ P(y'|x',s) \end{array} \right\} P(s).$$

Regarding the lower bounds of the $PN$ and the $PNS$, the conditional risk differences take the same signs in all the strata (stage 1: $-0.076$; stage 2: $-0.179$; stage 3: $-0.257$), which indicates that the proposed lower bound is equal to Tian-Pearl lower bound.

On the other hand, regarding the upper bounds of the $PN$ and the $PNS$, in stage 1 and stage 2, the signs of $\mathrm{P}(y|x,s) - \mathrm{P}(y'|x',s)$ are the same (stage 1: $-0.742$; stage 2: $-0.361$), but are different from that in stage 3 (stage 3: $0.457$), which indicates that the proposed upper bound is smaller than Tian-Pearl upper bound.

By calculation, the proposed bounds of the $PN$ is $(0.000, 0.778)$, and Tian-Pearl bounds is $(0.000, 1.000)$, which shows that the proposed bounds can provide more information for judging probability of causation. In addition, the proposed bounds of the $PNS$ is $(0.000, 0.168)$, which is also narrower than Tian-Pearl bounds $(0.000, 0.237)$.

## 6 Conclusion

Probabilities of causation are widely used in epidemiology, artificial intelligence and public policy analysis. Therefore, bounding and identifying the probabilities of causation is an important problem. In many experimental and observational studies, usually not only the exposure and outcome variables are measured, but also some covariates and intermediate variables are measured. This information enables us to narrow the bounds of the probabilities of causation. In this paper, we defined conditional probabilities of causation, and used this definition to propose narrower bounds than Tian-Pearl bounds by stratifying on some measured covariates and intermediate variables. We also consider the identification of the probabilities of causation based on stratified analysis. Since covariate selection problems occur in this situation, we further compared different cases of covariate selection from the viewpoint of estimation accuracy. Finally, we gave simulation results and analyzed an empirical data by using our formulas. With these new results added to the framework of Tian and Pearl (2000a, 2000b), the probabilities of causation should find wider use in more and more areas that require the evaluation of causal effects.

## Appendix

Regarding the $PN$, when $Y \perp\!\!\!\perp T \mid \{X, S\}$ holds true, we can obtain

$$\frac{\mathrm{P}(y|x,s)\mathrm{P}(y'|x,s)}{\mathrm{P}(x,s)}\mathrm{P}(x,s)^2$$

$$-\sum_t \frac{\mathrm{P}(y|x,s,t)\mathrm{P}(y'|x,s,t)}{\mathrm{P}(x,s,t)}\mathrm{P}(x,s,t)^2 = 0.$$

In addition, since we can obtain

$$\mathrm{P}(x'|s) \sum_t \frac{\mathrm{P}(x|t,s)^2 \mathrm{P}(t|s)}{\mathrm{P}(x'|t,s)}$$

$$= \sum_t \mathrm{P}(x'|t,s)\mathrm{P}(t|s) \sum_t \frac{\mathrm{P}(x|t,s)^2 \mathrm{P}(t|s)}{\mathrm{P}(x'|t,s)} \geq \mathrm{P}(x|s)^2$$

by the Cauchy-Schwarz Inequality,

$$\frac{\mathrm{P}(y|x',s)\mathrm{P}(y'|x',s)}{\mathrm{P}(x',s)}\mathrm{P}(x,s)^2$$

$$-\sum_t \frac{\mathrm{P}(y|x',s,t)\mathrm{P}(y'|x',s,t)}{\mathrm{P}(x',s,t)}\mathrm{P}(x,s,t)^2$$

$$= \mathrm{P}(y|x',s)\mathrm{P}(y'|x',s)\mathrm{P}(s)$$

$$\times \left( \frac{\mathrm{P}(x|s)^2}{\mathrm{P}(x'|s)} - \sum_t \frac{\mathrm{P}(x|t,s)^2 \mathrm{P}(t|s)}{\mathrm{P}(x'|t,s)} \right) \leq 0.$$

Thus, by noting that $PN_s = PN_{s,t}$ holds true, we can obtain $a.var(\hat{PN}_s) \leq a.var(\hat{PN}_{s,t})$.

Next, when $X \perp\!\!\!\perp S \mid T$ holds true, by the variance basic formula, we can obtain

$$\frac{\mathrm{P}(y|x,t)\mathrm{P}(y'|x,t)}{\mathrm{P}(x,t)}\mathrm{P}(x,t)^2$$

$$-\sum_s \frac{\mathrm{P}(y|x,s,t)\mathrm{P}(y'|x,s,t)}{\mathrm{P}(x,s,t)}\mathrm{P}(x,s,t)^2$$

$$\geq \sum_s \frac{\mathrm{P}(y|x,s,t)\mathrm{P}(y'|x,s,t)}{\mathrm{P}(x,s,t)} (\mathrm{P}(x,t)\mathrm{P}(x,s,t)$$

$$-\mathrm{P}(x,s,t)^2) \geq 0$$

and

$$\frac{\mathrm{P}(y|x',t)\mathrm{P}(y'|x',t)}{\mathrm{P}(x',t)}\mathrm{P}(x,t)^2$$

$$-\sum_s \frac{\mathrm{P}(y|x',s,t)\mathrm{P}(y'|x',s,t)}{\mathrm{P}(x',s,t)}\mathrm{P}(x,s,t)^2$$

$$\geq \sum_s \frac{\mathrm{P}(y|x',s,t)\mathrm{P}(y'|x',s,t)}{\mathrm{P}(x',s,t)} (\mathrm{P}(x,t)\mathrm{P}(x,s,t)$$

$$-\mathrm{P}(x,s,t)^2) \geq 0.$$

Thus, by noting that $PN_t = PN_{s,t}$ holds true, we can obtain $a.var(\hat{PN}_{s,t}) \leq a.var(\hat{PN}_t)$.

Regarding the $PNS$, when $Y \perp\!\!\!\perp T \mid \{X, S\}$ holds true, we can obtain

$$a.var(\hat{PNS}_{t,s}) - a.var(\hat{PNS}_s)$$

$$= \sum_x \left\{ \sum_s \frac{\mathrm{P}(y|x,s)\mathrm{P}(y'|x,s)}{N} \frac{\mathrm{P}(s)}{\mathrm{P}(x|s)} \right.$$

$$\left. \times \left( \mathrm{P}(x|s) \sum_t \frac{\mathrm{P}(t|s)}{\mathrm{P}(x|t,s)} - 1 \right) \right\}.$$

Here, by the Cauchy-Schwarz Inequality,

$$\begin{aligned} &\mathrm{P}(x|s) \sum_t \frac{\mathrm{P}(t|s)}{\mathrm{P}(x|t,s)} \\ &= \sum_t \mathrm{P}(x|t,s)\mathrm{P}(t|s) \sum_t \frac{\mathrm{P}(t|s)}{\mathrm{P}(x|t,s)} \geq 1. \end{aligned}$$

Thus, we can obtain $a.var(P\hat{N}S_s) \leq a.var(P\hat{N}S_{t,s})$.

When $X \perp\!\!\!\perp S \mid T$ holds true, by comparing $a.var(P\hat{N}S_t)$ with $a.var(P\hat{N}S_{t,s})$, we can obtain

$$\begin{aligned} &a.var(P\hat{N}S_t) - a.var(P\hat{N}S_{t,s}) \\ &= \sum_x \left\{ \sum_t \frac{\mathrm{P}(t)}{N\mathrm{P}(x|t)} \Big( \mathrm{P}(y|x,t)\mathrm{P}(y'|x,t) \right. \\ &\quad \left. - \sum_s \mathrm{P}(y|x,s,t)\mathrm{P}(y'|x,s,t)\mathrm{P}(s|t) \Big) \right\}. \end{aligned}$$

By using the variance basic formula, we can obtain $a.var(P\hat{N}S_{s,t}) \leq a.var(P\hat{N}S_t)$.

## Acknowledgements

The authors would like to thank Judea Pearl of UCLA for his helpful discussion about the paper. This research was supported by the Ministry of Education, Culture, Sports, Science and Technology of Japan, the Sumitomo Foundation, the Murata Overseas Scholarship Foundation, the Kayamori Foundation of Information Science Advancement, the College Women's Association of Japan and the Japan Society for the Promotion of Science.

**REFERENCES**

Bowden, R. J., and Turkington, D. A. (1984). *Instrumental Variables*, Cambridge University Press.

Cai, Z. and Kuroki, M. (2005). Variance Estimators for three " Probabilities of Causation ", *Risk Analysis*, **25**, 1611-1620.

Greenland, S. (1987). Variance estimators for attributable fraction estimates consistent in both large strata and sparse data. *Statistics in Medicine*, **6**, 701-708.

Greenland, S. (2000). An introduction to instrumental variables for epidemiologists. *International Journal of Epidemiology*, **29**: 722-729.

Greenland, S. (2003). Quantifying biases in causal models: Classical confounding versus collider-stratification bias. *Epidemiology*, **14**, 300-306.

Greenland, S. (2004). Model-based Estimation of Relative Risks and Other Epidemiologic Measures in Studies of Common Outcomes and in Case-Control Studies. *American Journal of Epidemiology*, **160**, 301-305.

Greenland S., Pearl J. and Robins J. M. (1999). Causal diagrams for epidemiologic research. *Epidemiology*, **10**, 37-48.

Newman, S. C. (2001). *Biostatistical methods in epidemiology*, Wiley.

Pearl, J. (2000). *Causality: Models, Reasoning, and Inference*, Cambridge University Press.

Robins, J. M. (1989). The analysis of randomized and non-randomized AIDS treatment trials using a new approach to causal inference in longitudinal studies. *Health Service Research Methodology: A Focus on AIDS*. Eds: Sechrest L., Freeman H., Mulley A. Washington, D.C.: U.S. Public Health Service, National Center for Health Services Research., 113-159.

Robins, J. M. and Greenland, S. (1989). The probability of causation under a stochastic model for individual risk. *Biometrics*, **45**, 1125-1138.

Rosenbaum, P. and Rubin, D. (1983). The central role of propensity score in observational studies for causal effects. *Biometrika*, **70**, 41-55.

Rubin, D. B. (2005). Causal Inference Using Potential Outcomes: Design, Modeling, Decisions. *Journal of the American Statistical Association*, **100**, 322-331.

Tian, J. and Pearl, J. (2000a). Probabilities of causation: Bounds and identification. *Annals of Mathematics and Artificial Intelligence*, **28**, 287-313.

Tian, J. and Pearl, J. (2000b). Probabilities of causation: Bounds and identification. *Proceedings of 16th Conference on Uncertainty in Artificial Intelligence*, 589-598.

Tian, J. (2004). Identifying conditional causal Effects. *Proceeding of 20th Conference on Uncertainty in Artificial Intelligence*, 561-568.